\documentstyle[11pt]{article}

\def\beq{\begin{equation}}
\def\eeq{\end{equation}}
\def\barr{\begin{eqnarray}}
\def\earr{\end{eqnarray}}
\def\tr{{\rm tr}}
\def\Tr{{\rm Tr}}

\begin{document}

\title{Planar QED in Magnetic or Electric Solitonic Backgrounds\footnote{Talk  
at Solitons Conference, Kingston, Ontario, July 1997}}

\author{Gerald Dunne\\Physics Department\\
University of Connecticut\\
Storrs, CT 06269 USA\\}

\date{}

\maketitle

\begin{abstract}
We review the resolvent technique for computing the effective action in planar
QED. For static magnetic backgrounds the effective action yields (minus) the
effective energy of the fermions, while for electric backgrounds the imaginary
part of the effective action gives (half) the probability of
fermion-antifermion pair creation. For some special `solitonic' background
profiles, these effective actions can be computed exactly.
\end{abstract}

\vskip 1cm

Consider the parity-even part of the 1-loop effective action for fermions in
$2+1$ dimensions in an external background electromagnetic field:
\beq
W=-\frac{i}{2}\Tr\,\log\,\left\{m^2-\left[\gamma\cdot (p-e A)\right]^2\right\}
\label{action}
\eeq
Using a standard integral representation of the logarithm \cite{schwinger}
\beq
W=\frac{i}{2}\int_0^\infty {ds\over s}\, \Tr\,exp\left(
-is\left\{m^2-\left[\gamma\cdot (p-e A)\right]^2\right\}\right)
\label{proper}
\eeq
This representation, known as Schwinger's proper-time formalism, yields an
elegant solution for backgrounds with constant field strength $F_{\mu\nu}$,
because in this case the proper-time propagator, $\Tr
\,exp(-is\{m^2-\left[\gamma\cdot (p-e A)\right]^2\})$,
has a simple closed form, which arises because the spectrum of the Dirac
operator is known exactly, and is purely discrete; so it is easy to take the
trace over the spectrum. For more complicated backgrounds, the exact spectrum
is not (in general) known, and even if it is known, it is no longer purely
discrete; so a more sophisticated method is needed to perform the energy trace.
The resolvent method discussed here provides such a method, and has led to
exact results in certain special nonuniform backgrounds \cite{cangemi,dunne}
(see also \cite{chodos}). Specifically, consider a background field that is
independent of one of the coordinates, and for which the corresponding
component of $A_\mu$ is zero. Take, for example, a case where $A_0=0$ and the
remaining fields are independent of time. This is appropriate for representing
a static, but possibly spatially nonuniform, magnetic background. Then the
trace over $p_0$ separates out and we obtain
\barr
W_{\rm mag}&\equiv&-\frac{i}{2}\Tr\,\log\left\{{\cal D}-p_0^2\right\}
=-i\int {dp_0\over 2\pi} \tr\left({p_0^2\over {\cal D}-p_0^2}\right)\nonumber\\
&=&\frac{1}{4\pi i}\int_C d\lambda\, \tr\left({\lambda^{1/2}\over {\cal
D}-\lambda}\right)
=\frac{1}{2}\tr\, {\cal D}^{1/2}
\label{resolvent}
\earr
In the first line we performed a formal integration by parts, and in the last
line we used Seeley's definition \cite{seeley} of the power of an operator:
\beq
{\cal D}^r={1\over 2\pi i}\int_C {\lambda^r\over {\cal D}-\lambda}\, d\lambda
\label{power}
\eeq
Note the appearance in (\ref{resolvent}) of $1/({\cal D}-\lambda)$, the
``resolvent'' of the operator ${\cal D}$. In (\ref{resolvent}), $\Tr$ includes
the trace over $p_0$ while $\tr$ refers to the remaining traces: the Dirac
trace and the spatial momenta corresponding to the eigenvalues of the static
operator ${\cal D}$. Using a ``Dirac'' representation for the Gamma matrices,
$\gamma^0=\sigma^3$, $\gamma^1=i\sigma^1$, $\gamma^2=i\sigma^2$, ${\cal D}$ is
diagonal:
\beq
{\cal D}=m^2+(\vec{p}-e \vec{A})^2+e F_{12}\sigma^3
\label{dmag}
\eeq
Note that the relation $W_{\rm mag}=\frac{1}{2}\tr\, {\cal D}^{1/2}$ simply
expresses the fact that the static effective action is minus the effective
energy; {\it i.e.} just a trace over the positive eigenvalues of the
corresponding Dirac operator.

For a constant magnetic field we choose $A_\mu=(0,0,B x)$, in which case each
diagonal component of ${\cal D}$ in (\ref{dmag}) is a harmonic oscillator
Schr\"odinger operator, with frequency $2eB$. Thus (including the overall
Landau flux degeneracy factor $eBA/2\pi$)
\barr
W_{\rm mag}&=&\frac{1}{2}{eB A\over 2\pi}\left(\sum_{n=0}^\infty
\sqrt{m^2+2eBn}+ \sum_{n=0}^\infty \sqrt{m^2+2eB(n+1)}\right)\nonumber\\
&=& {A(2eB)^{3/2}\over 8\pi}\left(\frac{|m|}{\sqrt{2eB}}+2\zeta(-\frac{1}{2},1+
\frac{m^2}{2eB})\right)
\label{zeta}
\earr
Restoring $\hbar$ and $c$, the governing dimensionless ratio is ${eB\hbar\over
m^2 c^3}$, which is the ratio of the cyclotron energy scale $\hbar \omega_c$ to
the fermion rest energy $mc^2$.

This same expression can be obtained using the resolvent method. Since ${\cal
D}$ involves only ordinary (not partial) differential operators, it is a simple
matter to find the resolvent (just the product of the two independent parabolic
cylinder eigenfunctions, divided by their Wronskian, and traced):
\beq
\tr\left({1\over {\cal D}-\lambda}\right)=-{A\over
4\pi}\left\{\psi\left({m^2-\lambda\over 2eB}\right)+
\psi\left(1+{m^2-\lambda\over 2eB}\right)\right\}
\label{psi}
\eeq
The psi function [$\psi(z)=(d/dz)log\Gamma(z)$] has simple poles at the
negative integers, so one reproduces (\ref{zeta}) using the contour integral
expression in (\ref{resolvent}). It is also an instructive exercise to compare
with the more conventional Schwinger proper-time computation which yields an
integral representation of the zeta function appearing in (\ref{zeta}).

The resolvent technique is unnecessarily complicated in this constant field
case, when ${\cal D}$ has a simple discrete spectrum. The power of this
technique becomes apparent for nonuniform backgrounds, for which the spectrum
of ${\cal D}$ is no longer purely discrete.

However, we first note that this formalism generalizes in a straightforward
manner to background fields that are independent of one of the spatial
coordinates (say $x^1$), and also $A_1=0$. The argument proceeds as before
except for some (physically significant!) factors of $i$. This case is
appropriate for describing a background electric field which may be time
dependent.
\barr
W_{\rm elec}&=&-\frac{i}{2}\Tr\,\log\left\{{\cal D}+p_1^2\right\}
=i\int {dp_1\over 2\pi} \tr\left({p_1^2\over {\cal D}+p_1^2}\right)\nonumber\\
&=&-\frac{1}{4\pi}\int_C d\lambda\, \tr\left({\lambda^{1/2}\over {\cal
D}-\lambda}\right)
=-\frac{i}{2}\tr\, {\cal D}^{1/2}
\label{elec-res}
\earr
In the ``chiral'' representation, $\gamma^0=-\sigma^1$, $\gamma^1=i\sigma^3$,
$\gamma^2=i\sigma^2$, ${\cal D}$ is diagonal:
\beq
{\cal D}=m^2-(p_0-e A_0)^2+(p_2-eA_2)^2+ie F_{02}\sigma^3
\label{delec}
\eeq
Notice that the eigenvalues of ${\cal D}$ are no longer real, and so $W_{\rm
elec}$ is not purely real. Since the probability of {\it no}
(fermion-antifermion) pair creation is $|e^{i W}|^2=e^{-2 Im(W)}\approx
1-2Im(W)$, this means that the imaginary part of the effective action gives
(half) the probability of pair creation \cite{schwinger}:
\beq
P({\rm pair}\,\,\,{\rm creation})\approx 2 Im(W)
\label{pair}
\eeq

For a constant electric field take $A_\mu=(0,0,Et)$; then each diagonal
component of ${\cal D}$ in (\ref{delec}) is a harmonic oscillator Schr\"odinger
operator with imaginary frequency $2ieE$. Then the effective action is given by
a zeta function expression as in (\ref{zeta}), with $B$ replaced by $iE$. Using
Hurwitz' representation of the zeta function \cite{bateman}, the imaginary part
yields
\beq
Im(W_{\rm elec})=A{(eE)^{3/2}\over 8\pi^2}\sum_{n=1}^\infty
{e^{-nm^2\pi/(eE)}\over n^{3/2}}
\label{imag}
\eeq
In the constant electric field case the governing dimensionless ratio is ${m^2
c^3\over eE\hbar}$; essentially the ratio of the pair-creation threshold energy
$2mc^2$ to the work done $eE \hbar/mc$ accelerating a charge $e$ over a fermion
Compton wavelength.

It is difficult to go beyond these constant field results
(\ref{zeta},\ref{imag}). In the `derivative expansion'
\cite{aitchison,lee,shovkovy} one assumes that the fields are `slowly varying'
with respect to a uniform background and expands formally in powers of
derivatives. For special forms of the background, one can do better - one can
find an exact integral representation \cite{cangemi} for the effective action
for a static magnetic background with a ``solitonic'' profile:
\beq
B(x,y)=B\, sech^2(x/ \lambda)
\label{soliton}
\eeq
where $B$ is a constant magnitude and $\lambda$ is a length scale. This problem
is solvable because the spectrum of the corresponding Dirac operator is known
exactly, and because the resolvent approach provides an efficient method for
tracing over both the discrete and continuum parts of the spectrum
\cite{cangemi}. The dimensionless combination $eB\lambda^2$ describes how flat
or sharp is the profile. In particular, when $eB\lambda^2\to \infty$ one
recovers the uniform field case. Correspondingly, the exact integral
representation for the effective action can be expanded in an asymptotic
expansion \cite{cangemi,dunne}
\barr
  &&W_{\rm mag} =\label{doublesum}\\
&&\hskip -15pt - \frac{L m^3 \lambda}{8 \pi} \sum _{j=0}^\infty
  {1\over j!(2eB\lambda^2)^j} \sum_{k=1}^\infty
 \frac{(2k+j-1)!{\cal B}_{2k+2j}}{(2 k)!(2k+j-\frac{1}{2}) (2k+j-\frac{3}{2})}
  \left(\frac{2eB}{m^2} \right)^{2k+j}\nonumber
\earr
The $j=0$ term in this sum agrees with the constant field answer, while the
$j=1$ term agrees with the first order derivative expansion contribution
\cite{cangemi2}. The general expansion (\ref{doublesum}) constitutes an {\it
all-orders} derivative expansion answer. This result has recently been extended
to $QED_{3+1}$ in \cite{dunne2}. Also, an analytic continuation $B\to iE$
yields a similar expansion for a time-dependent (but spatially uniform)
electric field
\beq
\vec{E}(t)=\vec{E}\, sech^2(t/ \tau)
\eeq
Some physical consequences of particle production rates in such a background
will be discussed elsewhere.
\vskip 1cm

\noindent{\bf Acknowledgements:}
This work has been supported by the DOE grant DE-FG02-92ER40716.00.

\end{document}